\documentstyle[12pt]{article}
\newcommand{\beq}{\begin{equation}}
\newcommand{\eeq}{\end{equation}}
\newcommand{\beqn}{\begin{eqnarray}}
\newcommand{\eeqn}{\end{eqnarray}}
\newcommand{\eq}[1]{(\ref{#1})}
\newcommand{\bi}[1]{\bibitem{#1}}
\newcommand{\fr}[2]{\frac{#1}{#2}}
\newcommand{\dst}{&\displaystyle}

\newcommand{\al}{Z\alpha}

\newcommand{\e}{\mbox{${\bf e}$}}
\newcommand{\eps}{\varepsilon}
\newcommand{\g}{\mbox{\boldmath ${\gamma}$\unboldmath}}
\newcommand{\gv}{\mbox{\boldmath $\gamma$ \unboldmath}}
\newcommand{\k}{\mbox{${\bf k}$}}

\newcommand{\pv}{{\mathbf p}}
\newcommand{\Q}{\kappa}
\newcommand{\r}{{\mathbf r}}

\newcommand{\rp}{{\mathbf r \,'}}

\newcommand{\vt}{\mbox{\boldmath ${\theta}$\unboldmath}}
\newcommand{\val}{\mbox{\boldmath ${\alpha}$\unboldmath}}

\newcommand{\vx}{{\bf x}}
\newcommand{\vv}{{\bf v}}

\newcommand{\q}{{\bf q}}
\newcommand{\vq}{\mbox{${\bf Q}$}}
\newcommand{\vd}{\mbox{${\bf \Delta}$}}

\newcommand{\grad}{\mbox{\boldmath ${\nabla}$\unboldmath}}

\begin{document}
\begin{titlepage}

\begin{center}
{\Large \bf Budker Institute of Nuclear Physics}
\end{center}

\vspace{1cm}

\begin{flushright}
{\bf Budker INP 99-06\\
February 1, 1999 }
\end{flushright}

\vspace{1.0cm}

\begin{center}{\Large \bf Simple analytical representation for the
high-energy Delbr\"uck scattering amplitudes}
\end{center}
\vspace{1.0cm}

\begin{center}
{{\bf R.N. Lee, A.I. Milstein, V.M. Strakhovenko}\\
G.I. Budker Institute of Nuclear Physics, \\
 630090 Novosibirsk, Russia}
\end{center}
\vspace{3cm}

\begin{abstract}
Using a new representation for the quasiclassical Green function
of the Dirac equation in a Coulomb field, analytical expressions for
the high-energy small-angle Delbr\"uck scattering amplitudes are obtained
exactly in the parameter $\al$. Magnitudes of the amplitudes coincide
with the previous results. However, the structure of the expressions
obtained is much more simple, which considerably facilitates
numerical calculations.
\end{abstract}

\end{titlepage}

The coherent scattering of photons in the electric fields of atoms via
virtual electron-positron pairs (Delbr\"uck scattering \cite{D}) has
been intensively investigated both theoretically and experimentally
\cite{MShu}. From the theoretical point of view this process is
interesting since the higher orders of
the perturbation theory in the parameter $\al$ ($Z|e|$ is the nucleus
charge, $\alpha =\, e^2\, =1/137$ is the fine-structure constant, $e$
is the electron charge, $\hbar =c=1$) drastically change the
amplitudes as compared to the first (Born) approximation. Delbr\"uck
scattering at high photon energy $\omega \gg m$ ($m$ is the electron
mass) is the most favorable to compare the theoretical and
experimental results.  Recently, the most accurate Delbr\"uck
scattering cross section measurements have been performed in the
Budker Institute of Nuclear Physics \cite{Budker1} at photon energies
$140-450$ MeV and scattering angles $2.6-16.6$ mrad. The Delbr\"uck
scattering amplitudes exact in the parameter $\al$ at high photon
energies and small scattering angle were obtained in
\cite{CW1,CW2,CW3} by summing in a definite approximation the Feynman
diagrams with an arbitrary number of photons exchanged with a Coulomb
center. Another representation of these amplitudes was derived in
\cite{MS1,MS2} using the quasiclassical Green function of the Dirac
equation in a Coulomb field. Such a Green function in an arbitrary
spherically symmetrical external field was obtained in \cite{LM1,LM2}
where Delbr\"uck scattering in a screened Coulomb field was
investigated as well.

In recent papers \cite{LMS1,LMS2,LMS3} the new approach was
developed, which significantly simplifies the consideration of
high-energy processes in a spherically symmetrical external field. It
was used to obtain the exact in $\al$ photon splitting amplitudes.
These results are used in the data processing of the first successful
experiment on the high-energy photon splitting in the electric fields
of atoms recently performed in the Budker INP (the preliminary
experimental results are published in \cite{Budker}).

In the present paper the approach developed in \cite{LMS1,LMS2,LMS3}
is applied to the calculation of the high-energy small-angle
Delbr\"uck scattering amplitudes. Numerically, the new amplitudes
coincide with the previously known results. However, the numerical
calculations by means of the new formulas, having actually only double
integral, are significantly easier.

As shown in \cite{LM1}, it is convenient to express the Delbr\"uck
scattering amplitude via the Green function of the squared Dirac
equation $D(\r,\rp |\,\eps )$:
$$
D(\r_1,\r_2 |\,\eps )=\langle\r_1|1/(\hat{\cal
P}^2-m^2+i0)|\r_2\rangle\ ,
$$
where $\hat{\cal P}=\gamma^{0}(\eps+\al/r )-\gv\pv$, $\pv=-i\grad$ .

In terms of the function $D(\r,\rp|\eps)$ the amplitude of the
process reads (\cite{LM1}):

\beqn\label{amplitude}
M&=&i\alpha\int\limits d\r_1 d\r_2\exp [i(\k_1\r_1 -\k_2\r_2)]\int\limits
d\eps\times \\
&&
\times\mbox{Sp}\biggl[(2\e_2^{*}\pv_2-\hat e_2^{*}\hat k_2)
D(\r_2 ,\r_1 |\omega -\eps)\biggr]
\biggl[ (2\e_1\pv_1+\hat e_1\hat k_1) D(\r_1 ,\r_2 |-\eps)\biggr]+
\nonumber\\
&&
+2i\alpha\e_2^{*}\e_1\,\int\limits d\r\exp[ i(\k_1 -\k_2)\r
]\,\int\limits d\eps\, \mbox{Sp} D(\r ,\r |\eps) \; , \nonumber
\eeqn
where  $e_1,\,k_1$ ($e_2,\,k_2$) are the polarization vector and
4-momentum of the initial (final) photon,
$\pv_{1,2} =-i\grad_{1,2}$. By definition, the Delbr\"uck scattering
amplitude \eq{amplitude} should vanish at $Z=0$. Therefore, one
should subtract from the integrand in \eq{amplitude} its value at
$Z=0$. We perform this subtraction in the explicit form below.
At $\omega\gg m$  the main contribution to the cross section comes
from the momentum transfer region $\Delta\sim m$ corresponding to the
small scattering angles. Then we can neglect the last term in
\eq{amplitude} since it depends only on the momentum transfer
$\vd=\k_2-\k_1$, while the amplitude at $\omega\gg \Delta$ is
proportional to $\omega$ (see, e.g., \cite{MShu}).

According to the uncertainty relation, the lifetime of the virtual
electron-positron pair is $\tau\sim |\r_2 -\r_1 | \sim
\omega/(m^2+\Delta^2)$ and $\rho \sim
1/\Delta$ for the characteristic impact parameter. For
$\omega\gg\Delta\gg m^2/\omega$ the main contribution to the
integral in \eq{amplitude} is given by small angles between the
vectors $\r_2 $, $-\r_1 $,  and $\k_1$. Then the characteristic
angular momentum is $l \sim \omega\rho \sim \omega/\Delta\gg 1$ and
the quasiclassical approximation is valid.
Note that the screening should be taken into account only at
$\Delta\sim r_c^{-1}\ll m$, where $r_c$ is the screening radius (
$r_c\sim (m\alpha)^{-1} Z^{-1/3}$ in the Thomas-Fermi model). In the
present paper we consider the momentum transfer region
$\Delta\gg m^2/\omega,\, r_c^{-1}$, which gives the main contribution
to the total cross section of the process. We emphasize that the
contribution of higher orders of perturbation theory in $\al$ (Coulomb
corrections) is not affected by screening and the corresponding
expressions are valid for any $\Delta\ll \omega$. The modification of
the Born contribution at $\Delta\sim m^2/\omega$ was
studied in \cite{CW2}, the effects of screening were discussed in
detail in \cite{LM1}.

In papers \cite{LMS1,LMS2}, the convenient representation for the
quasiclassical Green function $D(\r_1 ,\r_2 |\eps)$ of the squared
Dirac equation in a Coulomb field was obtained. For small
angles between  $\r_2 $, $-\r_1 $ and the $z$ axis, we have

\beqn\label{D1}
\dst
D(\r_1,\r_2 |\,\eps )= \,\fr{i\Q}{8\pi^2 r_1r_2}
\mbox{e}^{i\Q (r_1+r_2)}
\int d\q\, \left[1+\al\fr{\val\q}{\Q q^2}\, \right]\times \\
\dst
\times
\exp{\left [i\Q\fr{q^2(r_1+r_2)}{2r_1r_2}+i\Q\q\,(\vt_1+\vt_2)\right]}
\left(\fr{4r_1r_2}{q^2}\right)^{iZ\alpha\lambda} \quad \nonumber
,
\eeqn
where $\val = \gamma^0{\g}$ , $\Q^2=\eps^2-m^2$ , $\lambda=
\eps/\Q$ , $\q$, $\vt_1$, $\vt_2$ are two-dimensional vectors  in the
$xy$ plane, $\vt_1=\r_{1\perp}/r_1$ , $\vt_2=\r_{2\perp}/r_2$.
Expression \eq{D1} contains only elementary functions, and the angles
$\vt_1$ ³ $\vt_2$ appear only in the factor
$\exp[i\q\,(\vt_1+\vt_2)]$.  Therefore, representation \eq{D1} for
the Green function is very convenient for calculations.

We direct the $z$ axis along  $\k_1$. The z component of the
polarization vector $\e_{2}$ can be eliminated owing to the
relation $\e_2\,\k_2=0$, which leads to $(\e_{2})_z=\, -\e_{2}\vd/
\omega\,$ . After that, within the small-angle approximation one can
neglect the difference between the vector  $(\e_{2})_{\perp}$ and the
polarization vector of a photon, propagating along the $z$-axis and
having the same helicity. So, the amplitudes of the process are
expressed via the transverse polarization vectors $\e$ and $\e^{*}$,
corresponding to the positive and negative helicities, respectively.
It is sufficient to calculate two amplitudes, for instance, $M_{++}$
and $M_{+-}$. Other amplitudes ($M_{--}$ and $M_{-+}$) can be
obtained by the substitution  $\e\leftrightarrow\e^*$.

Substituting \eq{D1} into \eq{amplitude}, we expand the
amplitudes at small angles when $d\r_1d\r_2\approx r_1^2 r_2^2 dr_1
dr_2 d\vt_1 d\vt_2$.  Taking the trace and
performing the elementary integration over the angles
$\vt_1$ and $\vt_2$, we obtain
\beq\label{M1}
M=-\fr{i\alpha}{\omega^2}
\int\limits_{0}^{\omega}\! d\eps\, \eps\Q
\int\limits_0^{\infty}\!\fr{dr_1}{r_1}\!
\int\limits_0^{\infty}\!\fr{dr_2}{r_2}
\int\!\!\int \fr{d\q_1\,d\q_2}{(2\pi)^2}
\left[\left(\fr{q_1}{q_2}\right)^{2iZ\alpha}\!\!-1\right]
\mbox{e}^{i\Phi} T\, .
\eeq
Here
\beq
\Phi=\fr{1}{2}\left[\left(\fr{1}{r_1}+\fr{1}{r_2}\right)\vq^2
+\fr{\eps-\Q}{\omega}\vq \vd + \q\vd
 -m^2(r_1+r_2)\right]\, ,
\eeq
the function $T$ for different polarizations
has the form
\beq\label{T1}
T_{++}=\fr{2\vq^2}{r_1r_2}-\fr{\omega^2}{2\eps\Q}
\left(\fr{1}{r_1}+\fr{1}{r_2}\right)
\left[\left(\fr{1}{r_1}+\fr{1}{r_2}\right)\vq^2-2i\right]
\, ,\quad T_{+-}=\fr{4}{r_1r_2}(\e\vq)^2\,
\eeq
and following notation is introduced $\Q=\omega-\eps$,
$\vq=\q_1+\q_2$ ³ $\q=\q_1-\q_2$.  When deriving \eq{M1} , we
integrate by parts over $\q_1,\, \q_2$ so that the integrand
contains the parameter $\al$ only in the factor
$[(q_1/q_2)^{2iZ\alpha}-1]$. Besides, we make the substitution
$r_{1,2}\to (\eps\Q/\omega) r_{1,2} $.  The expression $T_{++}$ in
\eq{T1} can be simplified if we pass temporarily in \eq{M1} from the
variables $r_1,\, r_2$ to $R=r_1 r_2/(r_1+r_2)$ ,
$t=r_1/r_2$ and integrate by parts over $R$ the term proportional to
$\vq^2$ in square brackets in \eq{T1} . As a result, we get
\beq\label{T1a}
T_{++}=\fr{2\vq^2}{r_1r_2}+\fr{\omega^2m^2}{2\eps\Q
r_1r_2} (r_1+r_2)^2
\eeq
Let us pass from the variables $\q_1,\, \q_2$ to $\q$ and $\vq$.
Then the integral over $\q$ acquires the form:
\beq
J=\int\fr{d\q}{\vq^2}\,
\left[\left(\fr{|\q+\vq|}{|\q-\vq|}\right)^{2iZ\alpha}\!\!-1\right]
\exp(-\fr{i}{2}\q\vd)\quad .
\eeq
As shown in \cite{LMS1}, this integral is equal to
\beq
J=\int\fr{d\q}{\vd^2}\,
\left[\left(\fr{|\q+\vd|}{|\q-\vd|}\right)^{2iZ\alpha}\!\!-1\right]
\exp(-\fr{i}{2}\q\vq)\quad .
\eeq
Using this representation and the parametrization
\beq
\exp(i\fr{\vq^2}{2r_1})=ir_1\int \fr{d\vx}{2\pi}\exp(-i\fr{r_1
\vx^2}{2}-i\vq\vx)\ ,
\eeq
where $\vx$ is a two-dimensional vector lying in the same plane as
$\vq$, we take the integrals in \eq{M1} first over
$r_1$, then over $\vq$ and over $r_2$.
Finally, we have
\beqn\label{M3}
\left\{
\begin{array}{c}
M_{++} \\M_{+-}
\end{array}\right\}
&=&-\fr{i\alpha m^2}{\pi^2\Delta^2\omega^2}
\int\limits_0^\omega\!\! d \eps
\int\!\! d\q
\left[\left(\fr{q_+}{q_-}\right)^{2iZ\alpha}\!\!-1\right]\times\nonumber\\
&&
\times \int\!\!\fr{ d\vx}{(\vx^2+m^2)^2
(\vv^2+m^2)^2}
\left\{\begin{array}{c}
m^2(\eps^2+\Q^2)+\omega^2\vx\vv \\
4\eps\Q(\e\vv)^2\end{array}\right\}\ ,
\eeqn
where $q_\pm=|\q_\pm|$, $\q_\pm=\q \pm \vd$, $\vv=\vx+\q/2+\vd\,
(\eps-\Q)/2\omega$.

For the further integration it is convenient to rewrite
the expression \eq{M3} in another form. Using the identities
\beqn
\fr{m^2}{(\vv^2+m^2)^2}&=&\fr{1}{2}\grad_\vv\,\fr{\vv}{\vv^2+m^2}=
\grad_\q\,\fr{\vv}{\vv^2+m^2}\ ,\nonumber\\
\fr{\vv}{(\vv^2+m^2)^2}&=&-\fr{1}{2}\grad_\vv\,\fr{1}{\vv^2+m^2}=
-\grad_\q\,\fr{1}{\vv^2+m^2}\nonumber
\eeqn
and integrating by parts over $\q$, we find
\beqn\label{M4}
\left\{
\begin{array}{c}
M_{++} \\M_{+-}
\end{array}\right\}
&=&\fr{2\alpha(\al) m^2}{\pi^2\Delta^2\omega^2}
\int\limits_0^\omega\!\! d \eps
\int\!\! d\q
\left(\fr{q_+}{q_-}\right)^{2iZ\alpha}
\int\!\!\fr{ d\vx}{(\vx^2+m^2)^2 (\vv^2+m^2)}\times\nonumber\\
&&\times\left(\fr{\q_+}{q_+^2}-\fr{\q_-}{q_-^2} \right)
\left\{\begin{array}{c}
\omega^2\vx-(\eps^2+\Q^2)\vv\\
4\eps\Q(\e\vv)\e\end{array}\right\}\ .
\eeqn
Applying the Feynman parametrization for the denominators, we take
the integral over $\vx$ and pass from the variable $\eps$
to $s=2\eps/\omega-1$. We obtain

\beqn\label{M5}
\left\{
\begin{array}{c}
M_{++} \\M_{+-}
\end{array}\right\}
&=&\fr{4\alpha(\al) m^2\omega}{\pi\Delta^2}
\int\limits_{-1}^1\!\! ds
\int\!\! d\q
\left(\fr{q_+}{q_-}\right)^{2iZ\alpha}
\int\limits_0^1\!\!
\fr{t dt}{[t(1-t)(\q+s\vd)^2+4m^2]^2}\times\nonumber\\
&&\times\left(\fr{\q_+}{q_+^2}-\fr{\q_-}{q_-^2} \right)
\left\{\begin{array}{c}
[t(1-s^2)-2](\q+s\vd)
\\
2t(1-s^2) (\e,\q+s\vd)\e
\end{array}\right\}\ .
\eeqn
We perform the integration over $\q$ by means of a trick used in
\cite{LMS2,LMS3}.
Let us multiply the integrand in \eq{M5} by
\beqn\label{delta}
1&\equiv&\int_{-1}^{1}dy\,\delta
\left(y-\fr{2\q\vd}{\q^2+\vd^2} \right)
\nonumber\\
&=&(\q^2+\vd^2)\int_{-1}^{1}\fr{dy}{|y|}
\delta((\q-\vd /y)^2 -\vd^2(1/y^2-1)) , \nonumber
\eeqn
change the order of integration over $\q$ and $y$,
and make the shift $\q\rightarrow \q+\vd/y$. After that the
integration over $\q$ can be done easily. By changing the variables
$y=\tanh (\tau-\tau_0)$, where
\beq
\tau_0=\fr{1}{2}\ln \left(\fr{B+(1+s)^2}{{B+(1-s)^2}}\right)\,,\
B=\fr{4m^2}{\Delta^2t(1-t)}\ ,
\eeq
we express the integral over $\tau$ in terms of two functions, the
same as in \cite{LMS3}:
\begin{eqnarray}
\dst
{\cal F}_1
 =a^2\int_0^\infty d\tau\frac{\cosh \,\tau\,\cos \left( 2Z\alpha\tau\right)
 }{
\left( \sinh ^2\tau+a^2\right) ^{3/2}}=\fr{2\pi
a^2}{\sinh (\pi\al)}\mbox{Im}\,P_{i\al}^\prime(2a^2-1)\,
, \\
\dst
{\cal F}_2=a^2\int_0^\infty d\tau\frac{\sinh \,\tau\,
\sin \left( 2Z\alpha\tau\right) }{
\left( \sinh ^2\tau+a^2\right) ^{3/2}}=-\fr{2\pi
a^2}{\sinh (\pi\al)}\mbox{Re}\,P_{i\al}^\prime(2a^2-1)
\, , \nonumber
\end{eqnarray}
where $a^2=4B/[(B+(1+s)^2)(B+(1-s)^2)]$, $P_\nu^\prime(x)$ is the
derivative of the Legendre function. Note that $a^2\leq 1$ for any
$s$ and $B>0$.

The final result for the Delbr\"uck scattering amplitudes reads:
\beqn\label{MF}
M_{++}&=&i\fr{\alpha(\al) \omega}{8m^2}
\int\limits_{0}^1\!\! ds
\int\limits_0^1\!\!
dt\, a^2 t [2-t(1-s^2)]\times\\
&&\times
\biggl[4sB\sin (2\al\tau_0){\cal
F}_1+[B^2-(s^2-1)^2]\cos(2\al\tau_0){\cal F}_2\biggr]
\nonumber\ ,\\
M_{+-}&=&i\fr{\alpha(\al) \omega(\e\vd)^2}{4m^2\Delta^2}
\int\limits_{0}^1\!\! ds
\int\limits_0^1\!\!
dt\, a^2 t (s^2-1)\biggl[
4sB(1-t)\sin (2\al\tau_0){\cal F}_1+\nonumber\\
&&+
[B^2(2-3t)+2B(s^2+1)(1-2t)-(s^2-1)^2t]\cos(2\al\tau_0){\cal F}_2\biggr]
\nonumber\ .
\eeqn

Let us derive now the asymptotics of the amplitudes \eq{MF}
at $\Delta\ll m$ and $\Delta\gg m$. In the case $\Delta\ll m$,
when $B\sim m^2/\Delta^2\gg 1$, $a^2\approx 4/B\ll 1$, and $\tau_0\sim
1/B\ll 1$ , the calculations are especially easy. Then the functions
${\cal F}_{1,2}$ have the asymptotic form:
$$
{\cal F}_1\approx 1\,
, \ {\cal F}_2=-2\al\, a^2 [\ln a+C+\mbox{Re}\, \psi(1+i\al)] \, ,
$$
where $C=0.577\ldots$ is the Euler constant, $\psi(x)=d\ln
\Gamma(x)/dx$. Substituting the asymptotics of the functions
${\cal F}_{1,2}$ into \eq{MF}, we find
\beqn\label{Asym1} \left\{
\begin{array}{c}
M_{++} \\M_{+-}
\end{array}\right\}
&=&i\fr{4\alpha(\al)^2 \omega}{m^2}
\int\limits_{0}^1\!\! ds
\int\limits_0^1\!\!
t\, dt\,\left[\fr{1}{2}\ln \fr{m^2}{t(1-t)\Delta^2}-C-\mbox{Re}\,
\psi(1+i\al)\right]\times\nonumber\\
&&\times
\left\{\begin{array}{c}
[2-t(1-s^2)] \\2(1-s^2)(3t-2)(\e\vd)^2/\Delta^2
\end{array}\right\}\ .
\eeqn
Taking here the trivial integrals, we obtain at
$m^2/\omega\ll\Delta\ll m$:
\beqn\label{Asym11}
M_{++}&=&i\fr{28\alpha(\al)^2\omega}{9m^2}[\ln(m/\Delta)+\fr{41}{42}-C-
\mbox{Re}\,\psi(1+i\al)]\, , \nonumber\\
M_{+-}&=&i\fr{4\alpha(\al)^2\omega(\e\vd)^2}{9m^2\Delta^2}\ .
\eeqn
To obtain the asymptotics at $\Delta\gg m$,
it is convenient to start from \eq{M5}. The main contribution to the
integral over $t$ comes from the region  $1-t\sim
m^2/\Delta^2\ll 1$. Taking in \eq{M5} the integral over $t$ in this
approximation, we have:
\beqn\label{Asym2} \left\{
\begin{array}{c}
M_{++} \\M_{+-}
\end{array}\right\}
&=&\fr{\alpha(\al)\omega}{\pi\Delta^2}
\int\limits_{-1}^1\!\! ds
\int\!\! \fr{d\q}{(\q+s\vd)^2}
\left(\fr{q_+}{q_-}\right)^{2iZ\alpha}
\times\nonumber\\
&&\times\left(\fr{\q_+}{q_+^2}-\fr{\q_-}{q_-^2} \right)
\left\{\begin{array}{c}
-(1+s^2)(\q+s\vd)
\\
2(1-s^2) (\e,\q+s\vd)\e
\end{array}\right\}\ .
\eeqn
Substituting \eq{delta} into this formula, we take the elementary
integral over $\q$, and then over $s$. After that we
perform the substitution $y=\tanh\tau$ and obtain at $\Delta\gg m$:
\beqn\label{Asym3}
M_{++}&=&i\fr{4\alpha(\al) \omega}{3\Delta^2}
\int\limits_{0}^\infty\!\! d\tau\,\sin(2\al\tau)\,
[4-3\tanh(\tau/2)-\tanh^3(\tau/2)]\,=\\
&=&i\fr{8\alpha\omega}{3\Delta^2}\left\{1-\fr{2\pi\al}{\sinh(2\pi\al)}
[1-(\al)^2]\right\}\, ,\nonumber \\
M_{+-}&=&i\fr{16\alpha(\al) \omega(\e\vd)^2}{\Delta^4}
\int\limits_{0}^\infty\!\! \fr{d\tau\,\sin(2\al\tau)}{\sinh^2\tau}
(\tau\,\coth\tau -1)= \nonumber \\
&=&i\fr{16\alpha(\al)^2 \omega(\e\vd)^2}{\Delta^4}
[1-\al\,\mbox{Im}\,\psi^{\prime}(1-i\al)] \, . \nonumber
\eeqn
The asymptotics \eq{Asym11} and \eq{Asym3} coincide with the results
of \cite{CW3,MS2}. Additionally, we checked numerically for
different $Z$ that the magnitudes of the amplitudes \eq{MF}, as
should be, coincide with the results of \cite{CW3,MS1,MS2} at
intermediate values of the momentum transfer $\Delta$.
The expression \eq{MF} for the Delbr\"uck scattering
amplitudes is a double integral, being essentially simpler than the
known representations \cite{CW3,MS1,MS2}. Therefore, the  formula
\eq{MF} is very convenient for numerical calculations.

The calculation of the Delbr\"uck scattering amplitudes, performed in
the present paper, demonstrate again the advantage of our approach in
the consideration of different high-energy QED processes in a Coulomb
field.

\end{document}